 \documentclass[final,5p,times,twocolumn]{elsarticle}

\usepackage{amssymb}
\usepackage{graphicx}

\usepackage[cmex10]{amsmath}
\usepackage{comment}
\usepackage[footnotesize]{subfigure}

\usepackage{url}
\RequirePackage{amsthm}
\RequirePackage{amssymb}

\journal{Chaos, Solitons and Fractals}

\begin{document}

\begin{frontmatter}



\title{Disentangling bipartite and core-periphery structure in financial networks}


\author{Paolo Barucca} 
\address{Scuola Normale Superiore, Piazza dei Cavalieri 7, 56126 Pisa, Italy}
\author{Fabrizio Lillo}
\address{Scuola Normale Superiore, Piazza dei Cavalieri 7, 56126 Pisa, Italy}
\address{QUANTLab, via Pietrasantina 123, 56122 Pisa, Italy}
\date{\today}

\begin{abstract}
A growing number of systems are represented as networks whose architecture conveys significant information and determines many of their properties. Examples of network architecture include modular, bipartite, and core-periphery structures. However inferring the network structure is a non trivial task and can depend sometimes on the chosen null model. Here we propose a method for classifying network structures
and ranking its nodes in a statistically well-grounded fashion. The method is based
on the use of Belief Propagation for learning through Entropy Maximization
on both the Stochastic Block Model (SBM) and the degree-corrected Stochastic Block Model (dcSBM). 
As a specific application we show how the combined use of the two ensembles -SBM
and dcSBM- allows to disentangle the bipartite and the core-periphery
structure in the case of the e-MID interbank network. Specifically we find that, taking into account the degree, this interbank network is better described by a bipartite structure, while using the SBM the core-periphery structure emerges only when data are aggregated for more than a week.
\end{abstract}


\end{frontmatter}


\section{Introduction}

Network theory has become a unifying framework for describing and
understanding complex systems across various disciplines, from biology
to finance \cite{albert,newman}. The general assumption is that once the system
is represented as a network then network's properties will be significative
for the description of the system. In particular the analysis of the
network representation can give insight on system dynamics and helps
identifying the crucial quantities that drive the system's evolution, for
instance the preferential attachment mechanism in citations network.
One important property of a network, investigated in the present paper, is its organizational structure 
that we define as a statistically significant division of a network
in subnets with distinct properties.  Examples include bipartite \cite{strogatz,larremore}, modular or community \cite{karrer}, and core-periphery structure \cite{borgatti,holme,boyd,rombach,zhang}.
In general one would like to understand which structure is best-suited
to describe it through statistical inference, i.e. performing a sort of model selection. 

However finding structures in networks in a robust way is generically difficult. The
main problem is that one may try so hard to identify a structure in
a network where none exist and manage to find an 'illusory' structured
solution that is hard to distinguish from a 'real' statistically significant
structure. This happens, for example, in greedy community detection algorithms that
manage to find 'modular' structure also in random graphs. This problem is now clear in community detection and it has been thoroughly described and investigated through a free-energy based method to identify statistically significant communities \cite{zhangmoore}. 

In general, the problem of statistical significance of a structure
is either dependent from the class of models of correlated networks
that is considered or from the cost-function, for instance modularity,
that is used to assign a score to a structure. In this
paper we analyze the first case and we focus on the differences between
the network structures that are found through the Stochastic Block
Model (SBM) and the degree-corrected Stochastic Block Model (dcSBM).
In other words, we study how degree-heterogeneity affects the inference
of a network structure focusing specifically on the core-periphery
and bipartite structures.
The procedure we use to learn the parameters that characterize
the network structure is Belief Propagation \cite{decelle}. We thoroughly
analyze the problem of inferring a structure with SBM in a heterogeneous
network, i.e. with a broad distribution of degrees. As explained in
the seminal work of \cite{karrer} and investigated in \cite{decelle} for the community
detection problem, in degree-heterogeneous graphs, a SBM prior tends to clusterize
the nodes of a network in communities with similar degree. In particular
in the present work we numerically and analytically study the emergence
of a degree-based core-periphery structure in heterogeneous networks
having a different 'hidden' structure. The analysis allows us
to characterize the difference between a purely degree-based core-periphery
structure in heterogenous networks and a 'hidden' block structure,
describing the relations between blocks of nodes. Finally we apply
the analysis to a real financial interbank network at different levels of temporal aggregations . These results complement those found recently by us with a different inference method \cite{barucca}.
This diversity of results in networks at different levels of aggregation demonstrates the utility of the joint use of the two ensembles.

The paper is organized as follows. In the next section we outline the connection between the SBM, dcSBM, the configuration model and the class of correlated random networks with hidden variables
introduced in \cite{caldarelli, bogun}. Then we introduce a parametrization of the dcSBM
that reduces to the SBM for a particular value of the parameter and
briefly describe the learning procedure we use for the simulations
on the SBM and the dcSBM. In the last part of the section we obtain
the log-likelihood between adjacency matrices extracted from different
classes of correlated random networks, check analytically the existence
of the core-periphery bias in heterogeneous networks, and we present the numerical results on the interpolating class
of models and exhibit the emergence of the heterogeneity-driven core-periphery
structure in the SBM learning. In light of the numerical simulations in the second section,
we analyze the results of the SBM and dcSBM learning procedures on
data of a real interbank network. In the last section
we discuss results and suggest further theoretical enquiries for comparing
different classes of random network models, in particular to extend
the analysis to the case of weighted multiplex networks.

\section{Models and inference}

\subsection{Stochastic Block Models}

Statistical inference of communities is based on a Bayesian approach
with a prior random model parametrized by a finite set of parameters
that are recursively updated through a learning procedure \cite{decelle}.
SBM and dcSBM random-networks ensembles can be regarded as specific
cases of correlated random network ensembles \cite{bogun}. In correlated
random networks, nodes are associated with variables quantifying a
given number $M$ of different properties and each property influences
the probability of linkage between nodes. In particular the link probability
in the unweighted adjacency matrix $A=\{a_{ij}\}$ can be written in the
form:

\begin{equation}
p(a_{ij}=1)=\underset{r=1}{\overset{M}{\prod}}f_{r}(x_{i}^{r},x_{j}^{r})
\end{equation}
where $f_{r}(x^{r},y^{r})$ is a two-variable function that defines
the pairwise interaction between the $r$-th property $x$ of a pair
of nodes. In some cases the closer the values of a given property
the higher the probability of linkage - assortative properties - and
conversely in other cases the closer the values the lower the probability
- disassortative properties. 

In SBM with $m$ blocks we have a single property ($M=1$), the group assignment
$\{g_{i}\}_{i=1}^{N}$ where $g_{i}$ is an integer between $1$ and
$m$, and a single function $f_{1}(x,y)=p_{xy}$, where $m$ is the
number of groups and $p_{ab}$ is the link probability between a node
belonging to group $a$ and a node belonging to group $b$. The link probabilities $p_{ab}$ define
the $m\times m$ {\it affinity matrix}. More explicitly the link probability
reads: 

\begin{equation}
p(a_{ij}=1)=f_{1}(x_{i}^{1},x_{j}^{1})=p_{g_{i}g_{j}}
\end{equation}
SBM can generate, infer, and learn networks with a realistic modular
structure with the major drawback of being unable to reproduce networks
with fat-tailed degree distribution since the degree distribution
for a SBM network with a finite number of communities is just a mixture of Poisson distributions. The growing body of evidence
showing the strong degree-heterogeneity, i.e. the presence of fat-tailed degree distribution, in real networks has required
many theoretical efforts to understand the dynamical causes of such
a stylized fact \cite{albert} and also to include this feature in static
models of networks \cite{caldarelli}. 

To address this issue, Ref.  \cite{karrer} introduced dcSBM, where nodes have an extra property, the parameter $\theta_{i}$, and an extra function $f_{2}(x,y)=xy$\footnote{or its normalized \cite{karrer} counterpart $f_{2}(x,y)=xy/(1+xy)$ },
so that the link probability in this case reads: 

\begin{equation}
p(a_{ij}=1)=f_{1}(x_{i}^{1},x_{j}^{1})f_{2}(x_{i}^{2},x_{j}^{2})=p_{g_{i}g_{j}}\theta_{i}\theta_{j}
\end{equation}
In the case of no modular structure, that is when the affinity matrix
$p_{ab}$ is constant (i.e. there is only one block), we end up with the well-known configuration model \cite{newman}, also called hidden-variable  \cite{bogun} or fitness model \cite{caldarelli}, where the parameters $\{\theta_{i}\}$ control the degree of the nodes: the higher is $\theta_{i}$, the more probable is that node $i$ will have a high degree. For this reason  $\{\theta_{i}\}$ are called degree-corrections. Thus dcSBM can be regarded as a random configuration model with a modular structure. The major advance given by the dcBSM is the ability to take into account
both degree distribution and community structure,
so that statistical inference can be performed consistently on arbitrary
real networks, possibly scale-free networks with strong heterogeneity.

The subsequent implementation of a fast learning procedure \cite{decelle}
through a Belief-Propagation inference step has been a major accomplishment
in the community detection problem. In fact, while spectral algorithms
remain highly competitive because of the computational efficiency
of sparse linear algebra, these recent advances have improved the scalability 
of statistical inference with respect to the past \cite{decelle,ball}. Statistical inference starts from the full probability of an adjacency
matrix $A$ conditioned on the values of the ensemble parameters:

\begin{equation}\label{eq:prob}
P(A|p,\theta,g)=\prod_{i<j}(\theta_{i}\theta_{j}p_{g_{i}g_{j}})^{a_{ij}}(1-\theta_{i}\theta_{j}p_{g_{i}g_{j}})^{1-a_{ij}}
\end{equation}

Given an assignment $g$, maximizing this probability with respect to the degree-corrections
$\theta_{i}$ and to the elements of the affinity matrix $p_{ab}$
leads to $\theta_{i}=k_{i}/\kappa_{g_{i}}$ and $p_{ab}=m_{ab}/N$,
where $\kappa_{a}$ is the total degree of the nodes in block $a$,
$m_{ab}$ is the number of links between block $a$ and block $b$.

The full learning algorithm consists in: taking an initial guess for
the parameters $p$, $\theta$, and the block sizes, compute with Belief-Propagation the optimal assignment\footnote{This is obtained by maximizing the marginal assignment probabilities of single nodes}
for the given parameters, then compute the optimal parameters for
the found assignment and repeat the procedure till convergence \cite{decelle}. 

For sparse networks there is a regime in which statistical inference
methods can detect communities, while standard spectral algorithms
cannot \cite{decelle}. It has been showed that for sparse networks generated by the
SBM, spectral properties of the non-backtracking matrix are much better than those of the adjacency matrix
and its relatives \cite{krzakala}. In fact,
this method is asymptotically optimal, in the sense that it detects communities
all the way down to the detectability limit. The
spectrum of the non-backtracking matrix has been computed for some common benchmarks for community
detection in real-world networks, showing that the real eigenvalues
are a good guide to the number of communities and the correct labelling
of the vertices. Recently \cite{martin} the non-backtracking matrix has been
used also for defining a centrality measure, whose relation with core-periphery
is studied in \cite{inprep} 

\subsection{Comparing different ensembles}

The aim of this study is to understand the role of degree heterogeneity
in statistical inference from an analytical, numerical and experimental
point of view, thus in the following we systematically compare the SBM and
the dcSBM ensemble in an ad-hoc framework. In particular in the present
paragraph we use specific assignments to describe how networks with
increasing degree heterogeneity can experience a fundamental change in the
log-likelihood landscape. 

For a given network $A$, the log-likelihood that it has been generated by a dcSBM model with affinity matrix $\{p_{ab}\}$, assignments $\{g_i\}$, and the set of degree-corrections $\{\theta_i\}$ reads (see Eq. \ref{eq:prob}):
\begin{eqnarray}\label{likelygen}
L(A|p,\theta,g)=\frac{1}{2}(\sum_{ij}a_{ij}\log(p_{g_ig_j}\theta_i\theta_j)  \nonumber\\
+\sum_{ij}(1-a_{ij})\log(1-p_{g_ig_j}\theta_i\theta_j))
\end{eqnarray}
We consider the case where the set of degree-corrections of the generative dcSBM are i.i.d variables $\theta_i\in \mathbb{R}^+$ distributed according to a density $\psi(\theta)$ with mean value equal to one and extracted independently from the assignment $\{g_{i}\}$ and the affinity matrix $p_{ab}$.
If we now indicate with $\langle \cdot \rangle$  the average over the generative dcSBM ensemble, we have that $\langle{a}_{ij}\rangle = p_{g_ig_j}\theta_i\theta_j$.
Thus the average log-likelihood is 
\begin{eqnarray}
\langle L(A|p,\theta,g)\rangle=\frac{1}{2}(\sum_{ij}p_{g_ig_j}\theta_i\theta_j\log(p_{g_ig_j}\theta_i\theta_j)  \nonumber\\
+\sum_{ij}(1-p_{g_ig_j}\theta_i\theta_j)\log(1-p_{g_ig_j}\theta_i\theta_j))
\end{eqnarray}

We now consider a generic $\psi(\theta)$ and we take the sparse case limit \cite{chung}, i.e. when the link
density $\tilde{p}\equiv m^{-2}\sum_{a,b}p_{ab}\sim 1/N$ and the maximum of degree-corrections diverges slower
than $\sqrt{N}$, and we average $\langle L(A|p,\theta,g)\rangle$ over the degree-corrections distribution $\psi(\theta)$ obtaining 
\begin{equation}\label{likely}
\begin{split}
L_{real}[p,g, \psi(\theta)]/N = \overline{\langle L(A|p,\theta,g)\rangle}/N \\
\simeq\frac{1}{2}\left(\sum_{ab}c_{ab}n_{a}n_{b}\log(c_{ab})+2\tilde{c}\overline{\theta \log(\theta)} -\tilde{c}-\tilde c \log N\right). 
\end{split}
\end{equation}
Here we set $c_{ab}=Np_{ab}$, $\tilde c=N\tilde p$, and $n_a$ is the fraction of nodes in block $a$ in the assignment $\{g_i\}$. We also introduced the distribution averaging notation $\overline{f(\theta)}=\int d\theta f(\theta)\psi(\theta)$. Eq.\ref{likely} is quite instructive since it explicitly separates the block-structure term and the degree-correction term. 

Now we sample networks from a dcSBM consisting of two blocks of nodes ($m=2$) 
with a given structure of the affinity matrix and set of degree corrections and we compute the log-likelihood that a SBM model, characterized by different parameters and assignments but no degree corrections, generated the sampled network. 
In order to gain insight we restrict to a bimodal distribution of the degree-corrections, $\psi(\theta)=\frac{1}{2}\delta(\theta-\theta_1)+\frac{1}{2}\delta(\theta-\theta_2)$, where we set $\theta_1=1+\Delta$ and $\theta_2=1-\Delta$, with $\Delta<1$.
This is a strong limitation to the analysis but it constitutes a useful toy model, also used in \cite{karrer}, to understand how heterogeneity affects statistical inference with the SBM. 
Eq.\ref{likely} admits two known limits: one where no degree-corrections is present ($\theta_i=1$ for all $i$) and the optimal SBM assignment is simply given by $\{g_i\}$; the second where the elements of the affinity matrix are all equal and no community structure is present. In this case degree-corrections induce a specific structure, that is a core-periphery structure. In fact, in this case $c_{ab}=\tilde c \theta_a \theta_b$ ($a,b \in \{1,2\}$) and thus $c_{11}>c_{12}>c_{22}$.

With this consideration in mind we take the average log-likelihood of two specific SBM assignments:  

{\bf Block structure SBM:} a SBM associated with the true assignment and the true affinity matrix but ignoring degree-heterogeneity, whose average log-likelihood is:

\begin{eqnarray}\label{likelybs}
L_{bs}[p,g, \psi(\theta)]/N = \overline{\langle L(p,g|p,\theta,g)\rangle}/N\nonumber \\
\simeq\frac{1}{2}\left(\sum_{ab}c_{ab}n_{a}n_{b}\log(c_{ab})-\tilde{c}-\tilde c \log N\right)
\end{eqnarray}

{\bf Degree-based SBM:} a SBM where the assignment $g'$ is constructed starting from the degree-corrections, i.e. the nodes with $\theta_i=1+\Delta$ are put in the first block, the core, and the rest in the second block, the periphery. Consequently the SBM affinity matrix becomes $p{}_{ab}=\frac{\tilde{c}}{N}\theta_{a}\theta_{b}$,

\begin{eqnarray}\label{likelydb}
L_{db}[p,g, \psi(\theta)]/N = \overline{\langle L(p',g'|p,\theta,g)\rangle}/N \nonumber\\
\simeq\frac{1}{2}\left(\tilde{c}\log(\tilde{c})+2\tilde{c}\sum_{a}n_{a}\theta_{a}\log(\theta_{a})-\tilde{c}-\tilde c \log N\right)
\end{eqnarray}

The \textit{block structure} SBM log-likelihood $L_{bs}$ only accounts for the block term in Eq.\ref{likely}, while the \textit{degree-based} SBM log-likelihood $L_{db}$ only accounts for the degree-correction term and has an additional term $\tilde{c}\log(\tilde{c})$.
Hence the difference between $L_{real}$ and $L_{db}$ is given by the block structure term, while the difference between $L_{real}$ and $L_{bs}$ is given by the degree-correction term. By definition both $L_{bs}$ and $L_{db}$ remain smaller than $L_{real}$, since they are misspecified models. 
For small degree heterogeneity $L_{bs}$ is greater than $L_{db}$ because the block
structure term is larger than the degree-correction term, while for large
heterogeneity the converse is true and the preferred assignment between
the two is the degree-based one. Of course this does not tell us the
structure of the optimal assignment in the intermediate region but it is a clear indication of
how the log-likelihood landscape is changing with
heterogeneity, and how assignments with a core-periphery structure 
start to have an higher log-likelihood as heterogeneity increases.

\subsection{Numerical simulations}

We generate dcSBM bipartite networks, $c_{12}>c_{11}=c_{22}$, with a bimodal distribution of the degree-corrections, $\psi(\theta)=\frac{1}{2}\delta(\theta-\theta_1)+\frac{1}{2}\delta(\theta-\theta_2)$, with $\theta_1=1+\Delta$ and $\theta_2=1-\Delta$. In particular the affinity matrix reads:
\begin{equation}
p=\frac{1}{N}\left(\begin{array}{cc}
c & cr\\
cr & c
\end{array}\right)
\end{equation}
with $r=5$ and $c=2$ and $n_1=n_2=0.5$. Since $r>1$ the network organization is bipartite rather than core-periphery. We consider rather small networks of $N=80$ nodes since we want to understand the effects of heterogeneity on network sizes comparable with the real e-MID interbank networks investigated in Section 3, where the size ranges from 40 to 100 nodes. 

For each $\Delta$ we simulate a large number of samples, typically ranging from $100$ to $2,000$ and, by using Eq. \ref{likelygen}, we compute the log-likelihood according to different models. For each of them we average the log-likelihood across the samples, obtaining an ensemble average. Specifically we compute: (i) the log-likelihood of the generative model $L_{real}$, (ii) the log-likelihood of the \textit{block structure} SBM, $L_{bs}$, (iii) the log-likelihood of the \textit{degree-based} SBM $L_{db}$, and  (iv) the log-likelihood of the optimal assignment and optimal parameters, $n^*_a$ and $c^*_{ab}$ found with the Belief-Propagation learning procedure $L_{bp}$. Note that for the first three we use the exact expression of the log-likelihood and not the approximated expression in Eqs. \ref{likely}, \ref{likelybs}, and \ref{likelydb}.

\begin{figure}[t]\label{loglikes}
\includegraphics[width=80mm]{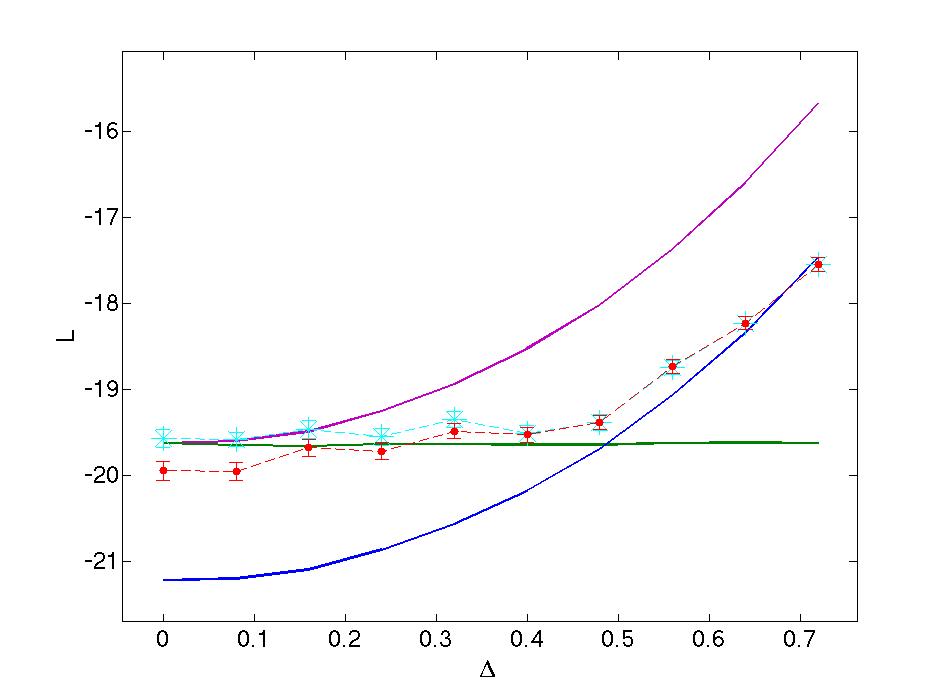}
\caption{The exact log-likelihood of the generative model $L_{real}$ (pink), the log-likelihood of the \textit{block structure} SBM (green), $L_{bs}$, the log-likelihood of the \textit{degree-based} SBM $L_{db}$ (blue), and the log-likelihood of the optimal assignment and optimal parameters, $n^*_a$ and $c^*_{ab}$ found with the Belief-Propagation learning procedure $L_{bp}$ (red). The turquoise line is the average log-likelihood of the optimal solution found with the BP learning algorithm restricted to the solutions with a core-size fraction between $0.4$ and $0.6$. Solid lines are averaged over 2000 samples while dashed lines are obtained from 100 samples. Error-bars lengths equal the standard deviations of the mean log-likelihoods, and are negligible for solid lines.}
\end{figure}

\begin{figure}[t]\label{structfracbim}
\includegraphics[width=80mm]{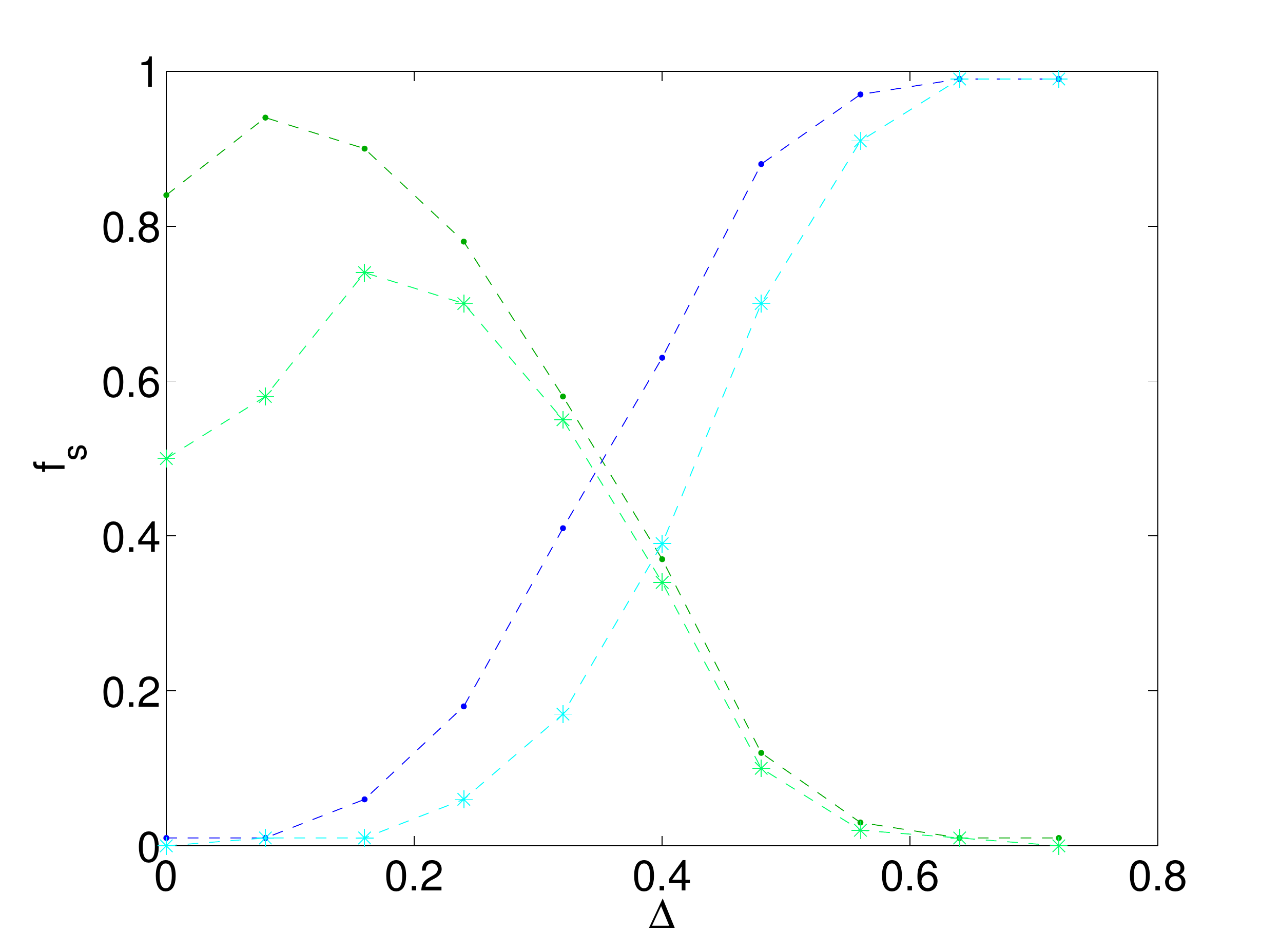}
\caption{The fraction of times BP learning algorithm identifies a bipartite (green dots) or a core-periphery (blue dots) structure. The fraction is measured over 100 networks and for each of them the BP learning algorithm is repeated 100 times and the optimal solution, and structure, chosen is the one with the lowest free-energy. When $\Delta$ increases the probability of finding a core-periphery structure goes to $1$. To take into account the variability in the core-sizes identified by the algorithm we plotted the light green and the turquoise dashed lines that are, respectively, the fraction of times a bipartite and a core-periphery with a core-size fraction $n^*_1$ between $0.4$ and $0.6$}
\end{figure}

The result is shown in Fig.1 where we clearly see the existence of the crossing-point between the $L_{bs}$ and $L_{db}$ (green and blue lines). This is consistent with the argument of the previous section, indicating, also in small size networks, that large degree heterogeneity  affects the log-likelihood landscape, favoring a misinterpretation of a bipartite network as a core-periphery structure. Fig.1 also shows that Belief-Propagation learning procedure for this size may get stuck in solutions with a likelihood $L_{bp}$ (red dashed line) lower than $L_{bs}$. The turquoise stars show the average $L_{bp}$ when the optimal solution found by the algorithm has a core size fraction $n^*_1\in[0.4,0.6]$, i.e.  close enough to the real fraction $0.5$. In this case we see that BP-algorithm finds optimal assignments associated with likelihoods respectively close to the block structure assignment for small $\Delta$ and to the degree-based assignment for large $\Delta$. We also checked that these optimal assignments for small and large values of $\Delta$ are indeed close to the block structure and the degree-based assignment respectively.

In Fig.2 we show the frequency of times BP-algorithm finds a bipartite or a core-periphery structure as a function of degree heterogeneity. We clearly see the emergence of the core-periphery structure due to degree-corrections and the disappearance of the bipartite solution, as also shown with a different analysis in \cite{karrer}.

Since in real-world applications degree-heterogeneity usually shows a broader distribution than a bimodal one, we also analyze the case of a power-law distribution for the degree-corrections, namely:
\begin{equation}
\psi(\theta) = \frac{\alpha-1}{\theta_{min}}(\frac{\theta}{\theta_{min}})^{-\alpha}
\end{equation}
where $\theta_{min}$ is fixed by the condition $\bar{\theta}=1$, so that the average degree is conserved varying $\alpha$. Note that the condition $p(a_{ij}=1)\le 1$ imposes that the tail exponent of the degree-correction distribution has to be bounded by the condition $\alpha>3$. The result is shown in Fig.3. Again we find that for increasing heterogeneity (small $\alpha$) BP finds more often a core-periphery structure while for small heterogeneity it finds more often and correctly a bipartite structure.

\begin{figure}[t]\label{structfrac}
\includegraphics[width=80mm]{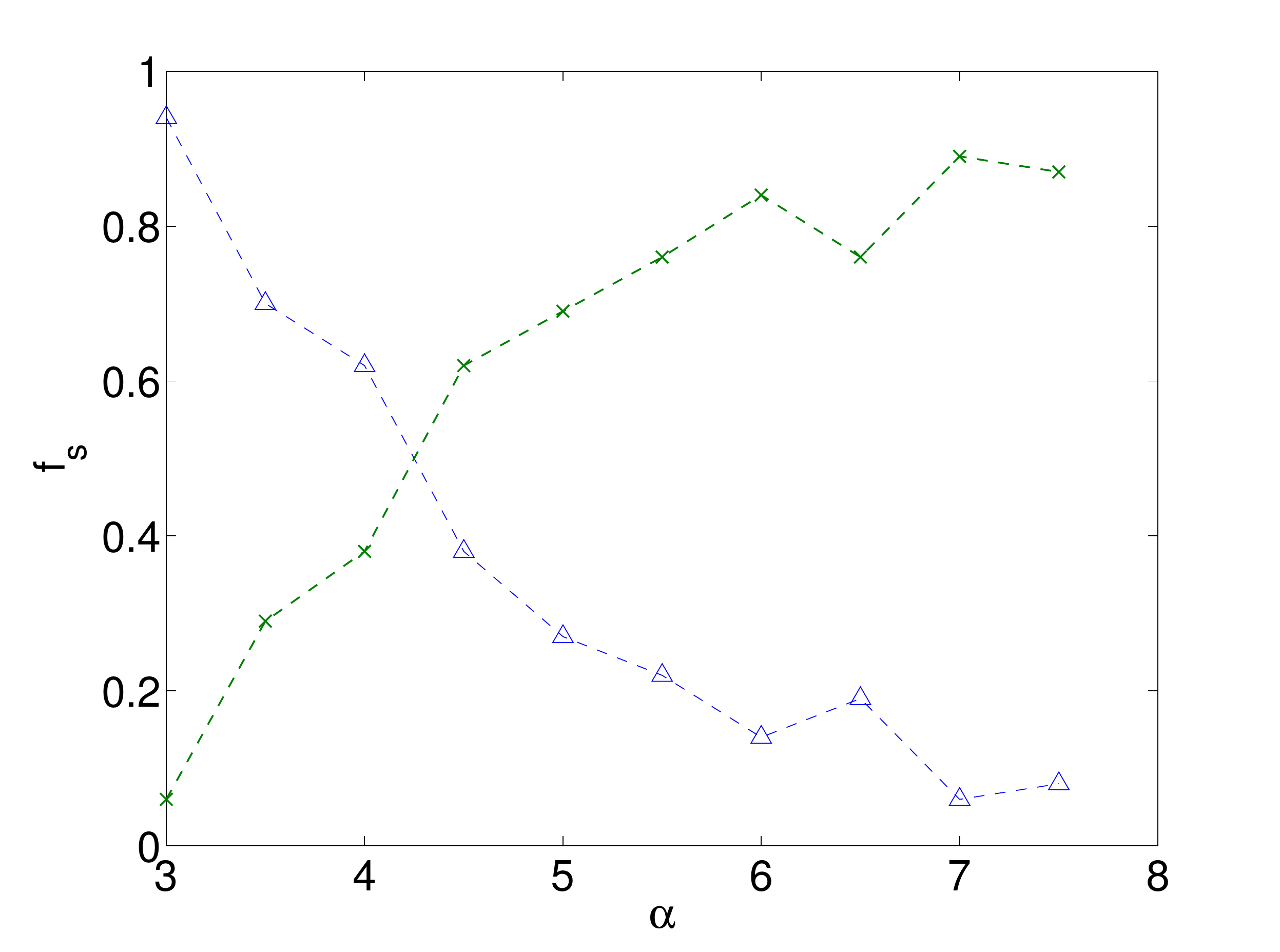}
\caption{The fraction of times BP learning algorithm identifies a bipartite (green crosses) or a core-periphery (blue triangles) structure. The fraction is measured over 100 networks and for each of them the BP learning algorithm is repeated 50 times and the optimal solution chosen is the one with the lowest free-energy. As the power-law exponent $\alpha$ decreases, from right to left, the degree-based core-periphery bias emerges more and more clearly.}
\end{figure}

\section{Large scale organization of interbank networks}

Interbank markets are a fundamental infrastructure of modern economies. They allow banks to lend and borrow money and therefore to finance themselves and, as a consequence, the whole economy. In the recent financial crisis the role of interbank markets as a monitoring system and its reaction to the harsh conditions of the economy have been deeply explored. Interbank markets are naturally represented as directed and weighted networks (or even multiplex networks \cite{bargigli}), where banks are the nodes and a credit relation is represented by a link.  Analysis on different interbank networks have agreed on several Óstylized factÓ or statistical regularities commonly observed: very low connectivity, an heterogeneous degree distribution, low average distance between nodes, disassortative mixing, small clustering, and an heterogenous level of reciprocity \cite{Boss,iori2008network,frickelux2,cont2011network}.

When considering the large scale organization of the network, it is often considered a core-periphery structure \cite{fricke,lelyveld}. However in a recent paper \cite{barucca} we revisit this finding by inferring a SBM on the e-MID interbank network (see below for more details) using a Markov Chain Monte Carlo (MCMC) approach \cite{tiago}. By considering a directed and weighted version of the network we find that for most aggregation time scales this interbank network is better described by a bipartite rather than by a core-periphery structure. Interestingly after Long Term Refinancing Operation (LTRO), one of the most important exceptional measure of ECB in the middle of sovereign debt crisis, the e-MID market is better described by a random structure, and only  two years after the LTRO it regained its bipartite structure. 

In this paper we perform a similar analysis to the one in \cite{barucca}, but (i) we use Belief Propagation rather than MCMC for inference and (ii) we consider  the undirected and unweighted version of the network. The last choice is motivated by the interest in the purely topological structure of the network and to identify better the transition from bipartite to core-periphery structure discussed in the previous section. In fact, also as shown in the Appendix of \cite{barucca}, the undirected and unweighted version of the interbank network is more frequently described by a core-periphery than when weights and directions are taken into account (at least for moderately large aggregation time scales).  

\subsection{The e-MID interbank network}

In this Section we investigate a real interbank time-evolving network and demonstrate how degree-bias is present in real-data analysis. We focus on the Italian electronic market for interbank deposits (e-MID) that 
is a platform for trading of unsecured money-market deposits operating in Milan
through e-MID SpA. The day-by-day network of overnight debt trading is constructed
from the list of transactions between banks where a bank, the giver, lends money
overnight, to another bank, the taker, that settles the debt the day after. We describe e-MID 
overnight network with the unweighted adjacency matrix $A(t)$. A generic element $a_{ij}(t)$ is $1$ 
if the bank $i$ lends money to bank $j$ the working day $t$ and bank $j$ settles the debt the working day 
after. The e-MID network has been thoroughly studied to understand bank liquidity management, as for instance in \cite{demasi, iori}. In particular here we analyze data from the $2^{nd}$ of January 2014 to the 
$31^{st}$ of December 2014. 

We want to investigate whether the interbank network is better described by a core-periphery or a bipartite network. To this end we perform SBM and dcSBM learning with belief-propagation algorithm for statistical inference and analyze its results varying 
the levels of aggregation. The economic reasons for our analysis are the following.  A core-periphery structure indicates the existence of a set of intermediary banks, the core, and a set of client-banks, the periphery, in need for the intermediation of the core. So core-periphery corresponds to a intermediated market for the e-MID network. On the other hand a bipartite structure, that a-posteriori in our case is found to be strongly directed \cite{barucca}, indicates the existence of a market without intermediaries where banks just trade following their liquidity needs and their preferences for the counterparts, i.e. displaying preferential trading.  We perform three kinds of analysis: firstly we analyze day-by-day 
structures, secondly we analyze month-by-month structures, where we aggregate the daily e-Mid network matrices 
over a month, and finally we investigate the dynamics of the cumulative matrix, obtained by increasing the level of aggregation. More specifically, for each day we consider the unweighted aggregated matrix $a^{(c)}_{ij}(t)$, 
where $a^{(c)}_{ij}(t)$ is $1$ if the bank $i$ has loaned money to bank $j$ overnight in any working day before day $t$. 

Since e-MID matrices can be rather small, i.e. number of banks performing transactions ranges from $40$ to $100$, and the learning procedure can get stuck in local minima of the log-likelihood, i.e. free energy, we repeat the fast BP learning procedure $100$ times and we also vary the initial condition over the affinity matrix, setting core-periphery or bipartite structure preserving the average degree of the given networks, in order to obtain robust results. 

As said above we consider the symmetrized version of the adjacency matrix $A$ obtained by taking the inclusive logical disjunction between $a_{ij}$ and $a_{ji}$ for each couple of banks.  We apply the existing learning algorithm available onÊ \url{http://mode_net.krzakala.org/} for symmetric matrices. The generalization to the weighted and directed case is certainly interesting and has been performed with the inference method of  \cite{tiago} in \cite{barucca}. Here we want to focus our attention on the role of degree heterogeneity in the detection of groups of undirected networks. 

We consider two global network metrics on the set of non isolated nodes. First we consider the density of links $\rho=\sum_{ij}a_{ij}/N^2 $. Fig.4 shows the density as a function of the number of days over which we aggregate the networks. It is clear that at one day scale the network is very sparse and its density increases significantly with aggregation scale. 
In order to measure the divergence from the random graph we compute the ratio $r_p=Var[k]/E[k]$ between the variance and the mean of the degree-distribution. A random graph has a Poissionian degree distribution and thus $r_p$ must equal one for a random graph. Large values of $r_p$ indicate a fat tailed degree distribution. Fig.5 shows $r_p$ as a function of the number of days over which we aggregate the networks. Also in this case a strong icrease is observed: at one day $r_p\simeq 1$, i.e. the degree distribution is close to a Poisson distribution, while when we aggregate over many days, a large value of $r_p$ is observed, signalling a strong degree heterogeneity. 

\begin{figure}[t]\label{figdens}
\includegraphics[width=80mm]{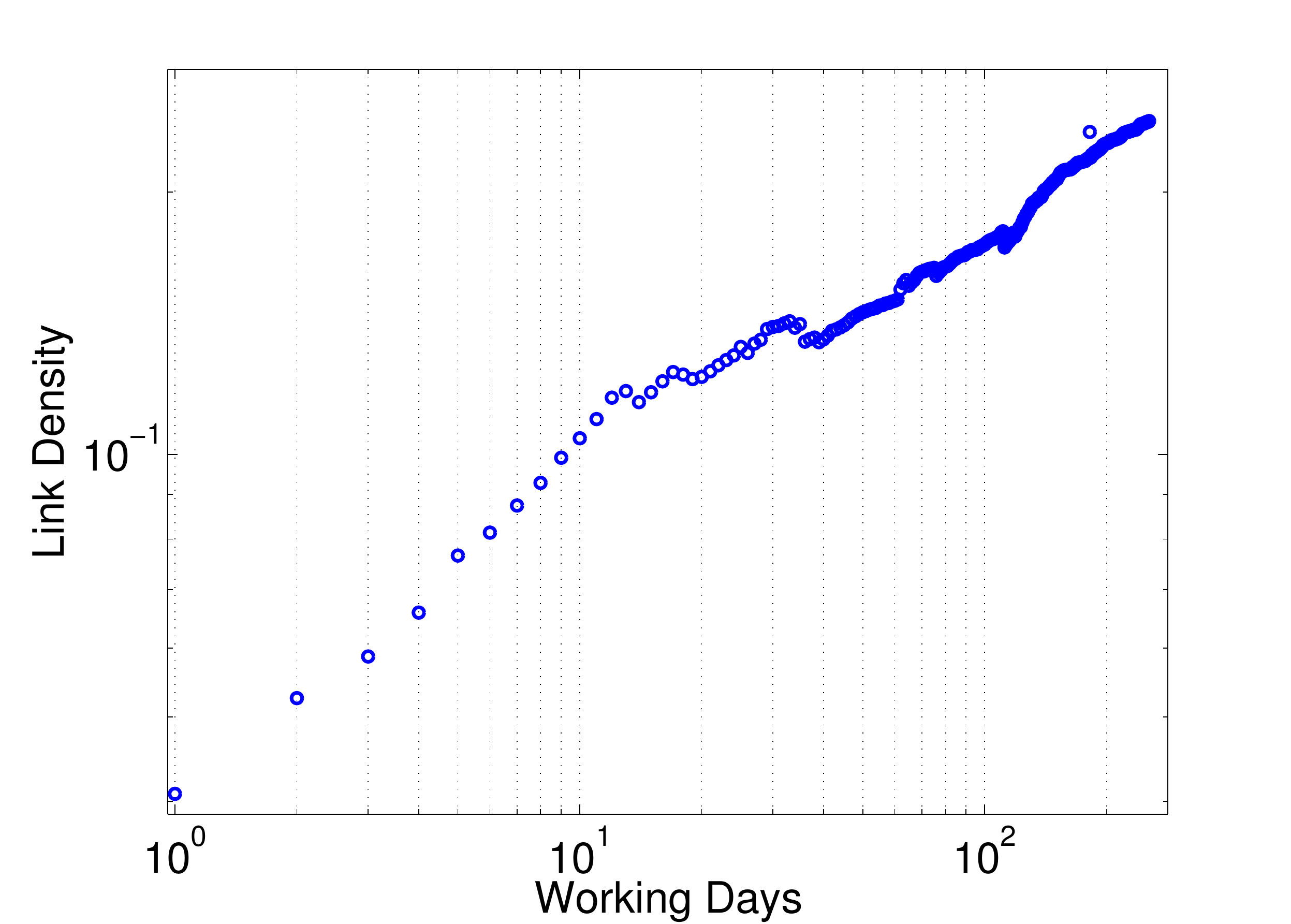}
\protect\caption{Density of links, $\rho=\sum_{ij}a_{ij}/N^2 $, as a function of the number of days over which we aggregate the interbank networks.}
\end{figure}

\begin{figure}[t]\label{figratio}
\includegraphics[width=80mm]{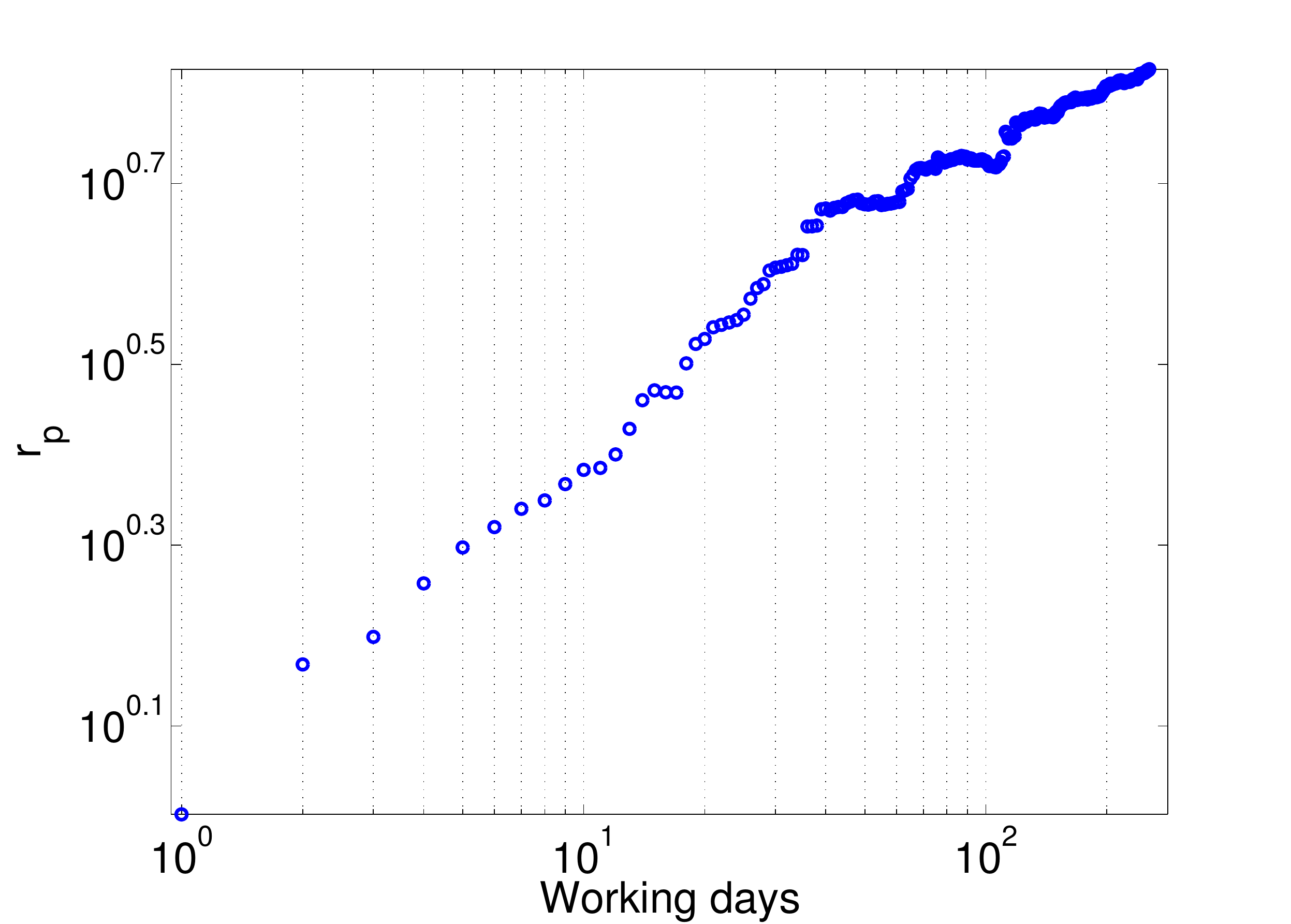}
\protect\caption{Ratio $r_p=Var[k]/E[k]$, where $k$ indicates degree, as a function of the number of days over which we aggregate the interbank networks. $r_p$ measures both degree heterogeneity and the divergence of the degree-distribution from a Poissonian distribution.}
\end{figure}

\paragraph{Daily structure}
Day-by-day analysis learning reveals mainly a bipartite structures with both ensembles. In fact we observe that in $65\%$ of the 2014 working days SBM finds a bipartite structure and in $91\%$ with dcSBM. Thus SBM and dcSBM agrees on the structure of the interbank network on a daily scale. A similar result is obtained with MCMC inference (see \cite{barucca}). This result is also consistent with the properties of the degree distribution. For daily matrices the maximum value of $r_p$ is $2.57$, thus the degree distribution is not exceedingly different from a Poissonian and SBM result is not strongly affected by degree-heterogeneity.  A clear indication of the bipartite structure of the network (and its directional properties) can be seen by considering the Laplacian matrix\footnote{The Laplacian matrix is $L=D-A$ where $D$ is the degree matrix (i.e. a diagonal matrix with nodes degrees in the non vanishing elements) and $A$ is the adjacency matrix.} with rows and columns sorted according
to the marginal probabilities computed at the learned values of the
parameters for the SBM  and dcSBM (panels (a) and (b) of Fig.6, for a specific day). Here, especially in the SBM case a clear bipartite and strongly directional structure (buyers on one side and seller on the other side) is observed. A strong bipartite day-by-day structure can be explained in a financial perspective considering that banks use e-MID market to balance their daily liquidity. Each day banks either have lacking or exceeding liquidity and thus they are either creditors or debtors. The analysis reveals that, in a given day, it is unlikely that a bank is both and thus in the e-MID market daily intermediaries are very rare.
\paragraph{Monthly structure}
We perform the same analysis on a monthly basis, i.e. considering the 12 unweighted adjacency matrices monthly aggregated. Here we obtain a different result, showing the importance of choosing the right null model and of taking into account degree heterogeneity. In fact, while SBM finds a core-periphery structure in 10 months (and hence a bipartite structure twice), dcSBM finds a bipartite structure 11 months (and a modular structure once). Panel (c) and (d) of Fig.6 shows the Laplacian matrix for a specific month with rows and columns sorted according to the marginal probabilities computed at the learned values of the parameters for the SBM  and dcSBM. The striking difference between the two cases shows that while the SBM finds a core-periphery structure, sorting the matrix according to dcSBM evidences a very clear bipartite and strongly directional structure also at monthly scale. The difference structure found by the two methods is due to the heterogeneity of degree. In fact, at monthly level of aggregation, the maximum value of $r_p$ is $4.12$, indicating fatter tail in the degree distribution. This is also consistent with our previous analysis: the more heterogeneous is the network, the more SBM learning is biased towards core-periphery structures. In conclusion, the core periphery structure observed in  interbank {\it binary} networks is in great part due to the fact that degree heterogeneity is not properly taken into account. 

\begin{figure}[t]\label{figlaplace}
\centering
\subfigure[SBM sorting on one day in  2014 ]{%
\includegraphics[scale=0.18]{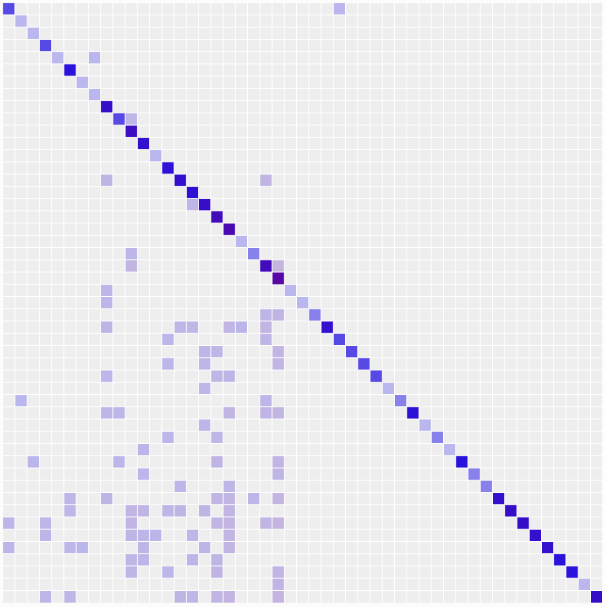}
\label{fig:subfigure1}}
\quad
\subfigure[dcSBM sorting on the same day of panel (a)]{%
\includegraphics[scale=0.18]{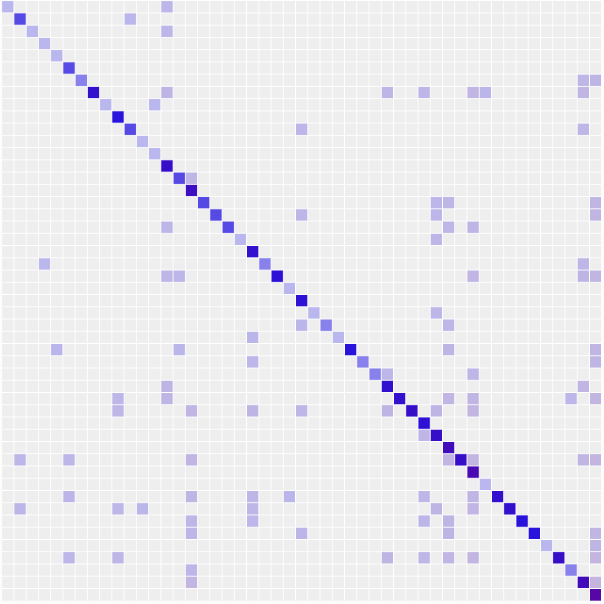}
\label{fig:subfigure2}}
\subfigure[SBM sorting on one aggregated month]{%
\includegraphics[scale=0.18]{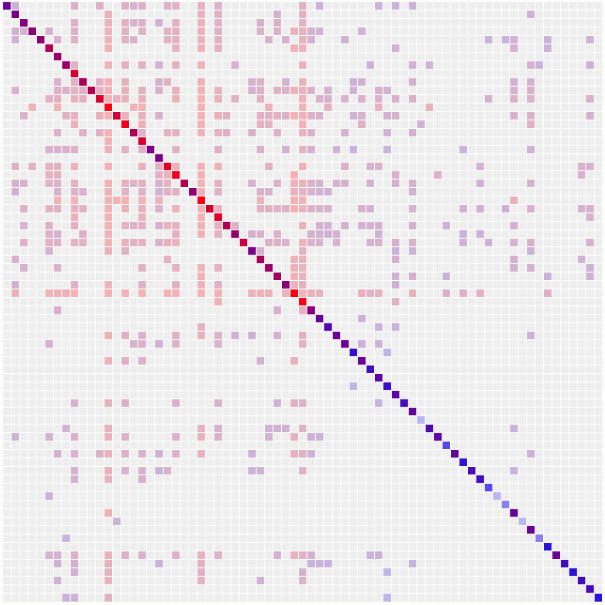}
\label{fig:subfigure3}}
\quad
\subfigure[dcSBM sorting on the same month of panel (c)]{%
\includegraphics[scale=0.18]{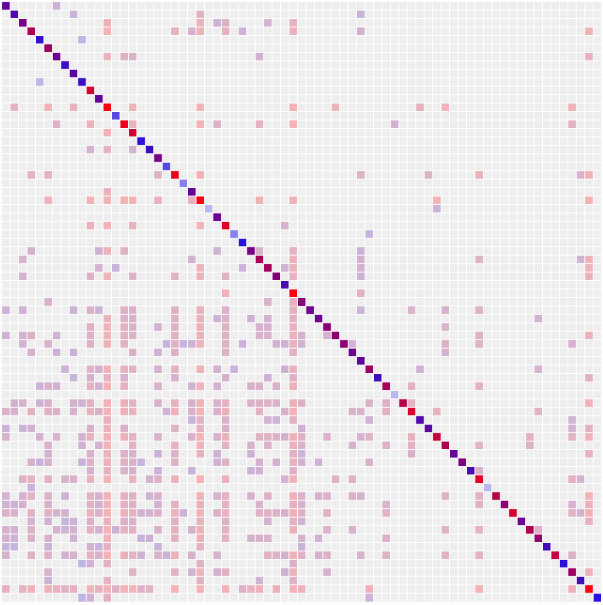}
\label{fig:subfigure4}}
\caption{Laplacian matrices of  e-MID interbank network sorted according
to the marginal probabilities computed at the learned values of the
parameters for the SBM {[}left{]} and for the dcSBM ensemble {[}right{]}.
Color scale of links is proportional to nodes degree. Both ensembles
give a ranking which is different from the one given by nodes degree
and dcSBM ranking highlights the directed and bipartite structure
of the network.}
\end{figure}

\paragraph{Aggregation-time dependence of structure learning}
Finally we investigate how time aggregation affects the structure of the network and the corresponding learning process. We want also to  check if results are stable across time-scales. Therefore we analyze the affinity matrices obtained by SBM and dcSBM on $A^{(c)}(t)$ as functions of the number of days in the aggregation. To avoid effects due to varying density with time scale, we normalize the affinity matrix so that the sum of the elements is one, i.e. $\hat p_{ij}=p_{ij}/\sum p_{ij}$.
In Fig.7 we show the behavior of the elements of the normalized affinity matrix for SBM as a function of the number of days over which we aggregate the interbank networks. We observe a clear transition. While for an aggregation of less than three days, SBM identifies a bipartite structure ($\hat p_{12}>\hat p_{11}>\hat p_{22}$), for larger aggregations SBM identifies a clear core-periphery structure ($\hat p_{11}>\hat p_{12}>\hat p_{22}$). Notice also that after ten days of aggregation the estimated parameters are pretty stable. 
The learning with dcSBM gives a completely different picture. At all time scales the bipartite structure is identified, since $\hat p_{12}>\hat p_{11}>\hat p_{22}$, suggesting that the division of banks in a set of creditors and a set of debtors persists. Also in this case the structure is quite stable, even if a decline of $\hat p_{12}$ is visible. As before, the different behavior in the figures can be explained by the fact that the increasing heterogeneity of degree with time scale (see Fig.8) is seen by the SBM as the emergence of a core-periphery structure, while the dcSBM, taking into account this heterogeneity, identifies a bipartite structure at all time scales. 

\begin{figure}[t]\label{figaffinitysbm}
\includegraphics[width=80mm]{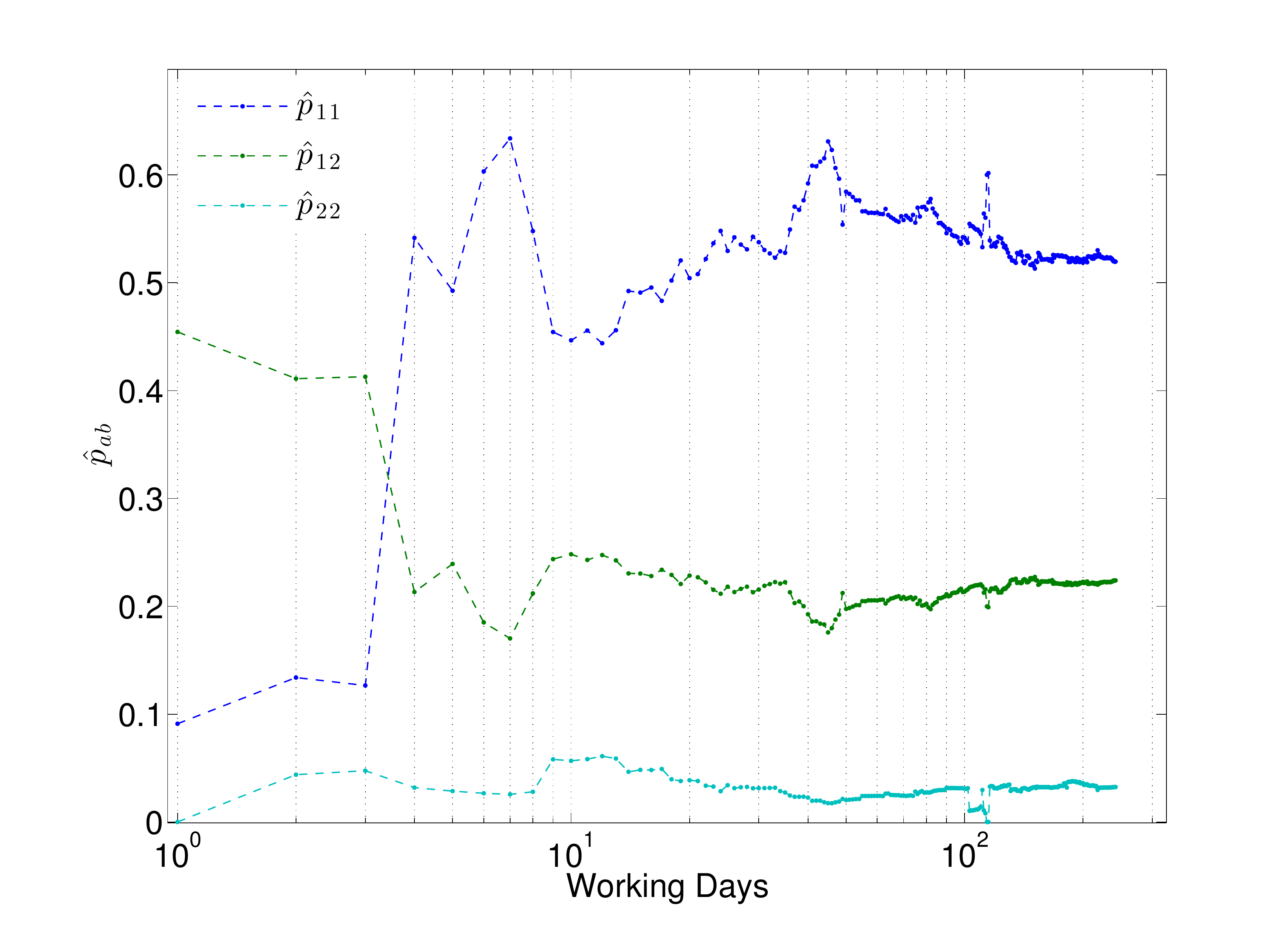}
\protect\caption{Elements of the normalized affinity matrix of the Stochastic Block Model on the aggregated matrix $A^{(c)}(t)$ as a function of the number of days over which we aggregate the interbank networks.}
\end{figure}

\begin{figure}[t]\label{figaffinitydcsbm}
\includegraphics[width=80mm]{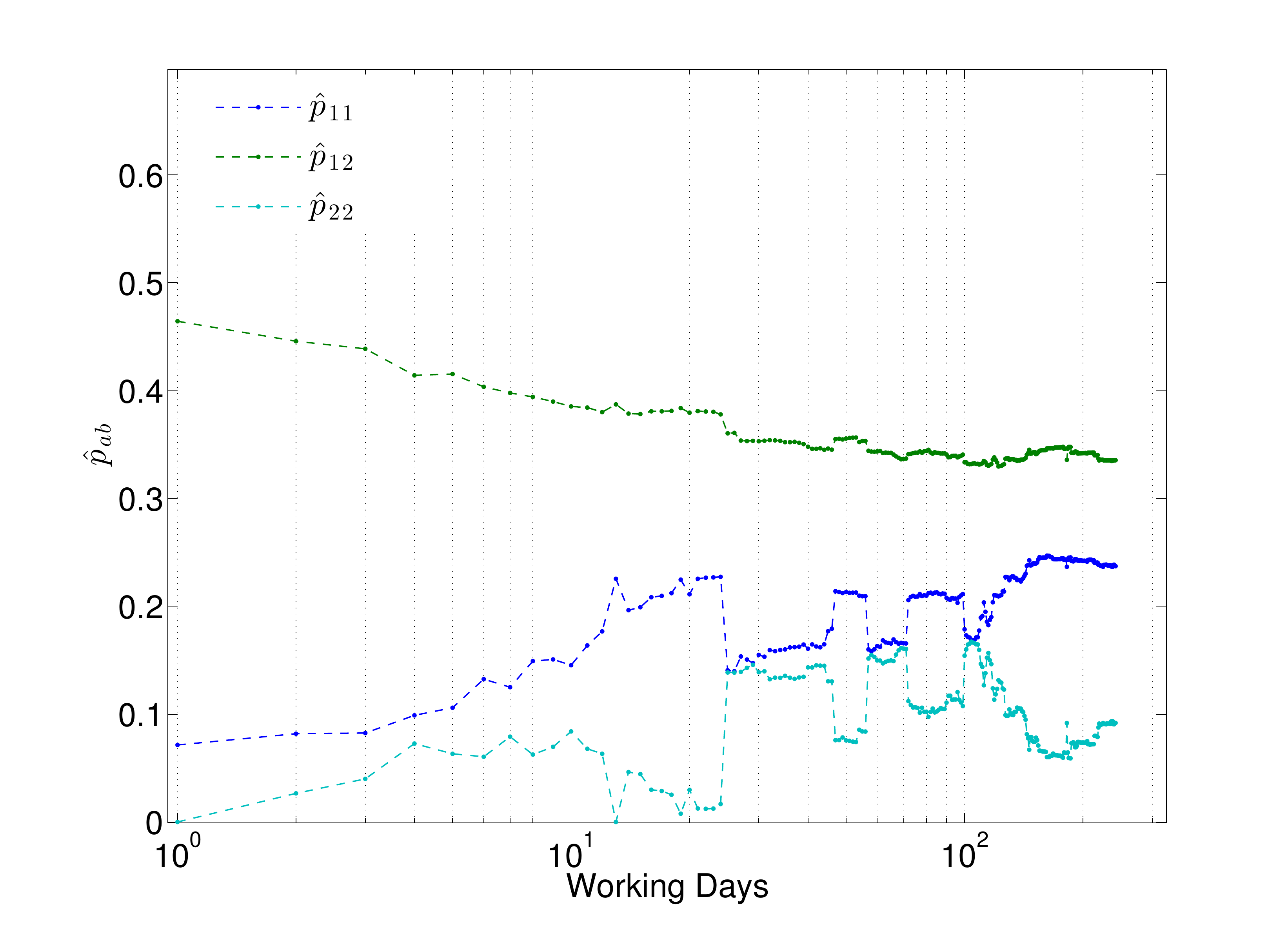}
\protect\caption{Elements of the normalized affinity matrix of the degree-corrected Stochastic Block Model on the aggregated matrix $A^{(c)}(t)$ as a function of the number of days over which we aggregate the interbank networks.}
\end{figure}

\section{Conclusions}
In the present work we analyzed the role of heterogeneity in structure learning of time-varying networks. The role of degree-heterogeneity has been widely investigated in the community detection context but its role in the recent core-periphery identification problem had not been previously outlined and analyzed in detail. We suggested a general framework to compare generative models that consists in evaluating the log-likelihood that a specific set of parameters of a given generative model actually generated a network sampled from another set of parameters of another generative model. This simple approach allowed us to understand the emergence of the core-periphery bias in SBM learning on dcSBM networks. 
The importance of this analysis is illustrated with the application on the e-MID interbank network. The comparison of heterogeneity, SBM and dcSBM learning reveals in this case a bipartite structure stable across time-scales of aggregation that is hidden by degree-heterogeneity that grows with aggregation. The bipartite structure in this interbank network corresponds to the absence of purely intermediary banks. 
In this work we used the available algorithm described in \cite{decelle}, in the unweighted case and although a framework for weighted networks has been introduced in \cite{karrer} still a general inference algorithm that accounts for strengths, degree heterogeneity, and multiple layers has not been established yet, though efforts are being made in this direction \cite{tiago3}, and it will be subject for future work. 

\section*{Acknowledgment}

Authors acknowledge partial support by the grant SNS13LILLB ''Systemic risk in financial markets across time scales". This work is supported by the European CommunityÕs H2020 Program under the scheme ÔINFRAIA-1-2014-2015: Research InfrastructuresÕ, grant agreement \#654024 ÔSoBigData: Social Mining \& Big Data EcosystemÕ (http://www.sobigdata.eu).

\end{document}